\documentclass[twocolumn,english,aps,prl,showpacs]{revtex4}
\usepackage{CJK}
\usepackage{graphicx}
\usepackage{color}
\usepackage{amsmath}
\usepackage{amssymb}
\usepackage{amsthm}
\usepackage{booktabs}
\usepackage[dvipdfm,colorlinks,urlcolor=blue,linkcolor=blue,anchorcolor=blue,citecolor=blue]{hyperref}

\graphicspath{{Graphs/}}
\input{tcilatex}
\begin{document}

\title{Effective phonon treatment of asymmetric interparticle interaction
potentials}
\author{Weichen Fu$^{1}$,Hong Zhao$^{1,2}$}
\email{zhaoh@xmu.edu.cn}

\affiliation{Department of Physics and Institute of Theoretical Physics and Astrophysics,
Xiamen University, Xiamen 361005, Fujian, China}
\affiliation{Collaborative Innovation Center of Chemistry for Energy Materials, Xiamen
University, Xiamen 361005, Fujian, China}

\date{\today }
\begin{abstract}
We propose an effective phonon treatment in one dimensional
momentum-conserved lattice system with asymmetric interparticle interaction
potentials. Our strategy is to divide the potential into two
segments by the zero-potential point, and then approximate them by piecewise harmonic potentials with effective force constants $\tilde{k}_L$ and $\tilde{k}_R$ respectively. The effective
phonons can then be well described by $\omega_c=\sqrt{2(\tilde{k}_L+\tilde{k}_R)}%
|sin(\frac{1}{2}aq)|$. The numerical verifications show that this treatment works very well. 
\end{abstract}

\pacs{xxx}
\maketitle

Effective phonon treatment(EPT) has significant importance in condensed matter physics
\cite%
{PhononsBook1990,AlabisoJSP1995,LepriPRE1998,GershgorinPRL2005,GershgorinPRE2007,BaowenEPL2006,BaowenEPL2007,HePRE2008,NianbeiPRL2010,Liusha14PRB}
Refs.\cite%
{AlabisoJSP1995,LepriPRE1998,GershgorinPRL2005,GershgorinPRE2007,BaowenEPL2006,BaowenEPL2007,HePRE2008,NianbeiPRL2010}%
) and it seems they do effective in many situations. In the past two decades, the heat
conduction in low-dimensional systems has attracted intensive studies\cite{Lepri03,Dhar08,Baowen12RMP}(see also references therein), 
and the EPT is applied also to this problem since phonons act as the predominant heat carrier. The
scaling behaviors of heat conduction have been successfully explained \cite%
{BaowenEPL2006,BaowenEPL2007,HePRE2008}, and the sound speed is predicted \cite{BaowenEPL2006,NianbeiPRL2010}.

\begin{figure}[tbph]
\centering
\includegraphics[width=1\columnwidth]{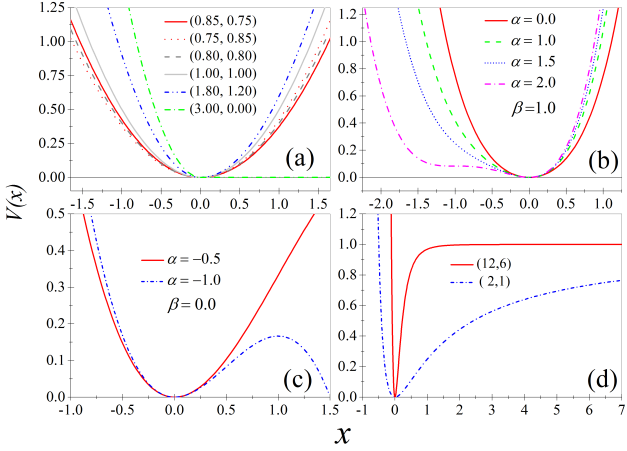}
\caption{(Color lines) (a) to (d) are plots of potential of piecewise model, FPU-$\protect\alpha $-$\protect\beta $ model, FPU-$\protect\alpha $
model, and L-J model, respectively.}
\label{fig_potential}
\end{figure}

Recently, Y. Zhang et al.\cite{ZhangyongarXiv2013} pointed out that while EPT works well in lattices with symmetric interaction potential,
significant divergence occurs in lattices with asymmetric interaction potential even in predicting the sound velocity. 
In the present paper, we present a simple but very effective treatment of effecive phonons. Our idea is
to divide the interaction potential into two segments by the minimum potential point. Then the
left and the right segments are equivalent to two segments of harmonic potential respectively. The piecewise harmonic potential is applied to derive the properties of the unharmonic system analysis. We will varify our idea by analytically and numerically using several typical one-dimentional lattice models, including the Fermi-Pasta-Ulam-$\alpha$-$\beta$ (FPU-$\alpha$-$\beta$) model \cite%
{FPU} and the Lennard-Jones (L-J) model.The Hamiltonian we study reads 
\begin{equation}  \label{eq-Hamiltonian}
H=\sum_i^N\big[\frac{p_i^2}{2}+V(x_i-x_{i-1})\big],
\end{equation}
where $p_i$ is the momentum and $x_i$ is the displacement from equilibrium
position for the $i$th particle, $N$ is the total number of particles that
equals system's size (the lattice constant is set be unit in our studies below), $V$ is the
interaction potential between nearest neighbour lattices. 
For the FPU-$\alpha$-$\beta$ lattice, $V$ is given as 
\begin{equation}  \label{eq-FPU-V}
V(x)=\frac{1}{2}x^2+\frac{\alpha}{3}x^3+\frac{\beta}{4}x^4,
\end{equation}
where $\alpha$ controls the degree of asymmetry (see Fig. \ref{fig_potential}%
(b) and (c)). While $\alpha=0$, $\beta\ne0$, it becomes symmetric FPU-$\beta$ model. On the contrary, if $\alpha\ne0$, $\beta=0$, it
appears as the FPU-$\alpha$ model (see Fig. \ref%
{fig_potential}(c)) \cite{FPU}. The potential of the L-J model is 
\begin{equation}
V(x)=(\frac{1}{1+x})^m-2(\frac{1}{1+x})^n+1,
\end{equation}
where the parameter set ($m$, $n$) control the degree of asymmetry (see Fig. %
\ref{fig_potential}(d)). This model has important practical implications
because it can well approximate the inter-particle interactions in many real
materials.

As the harmonic potential model is applied as the templet model in traditional EPTs, we also need a templet. It is the piecewise harmonic model defined as 
\begin{equation}  \label{eq-asy-harmonic-V}
V(x)=%
\begin{cases}
\frac{k_{L}}{2}x^2, & x<0 \\ 
\frac{k_{R}}{2}x^2, & x\ge0 \\ 
& 
\end{cases}%
,
\end{equation}
where $k_{L}$/$k_{R}$) is the left/right force constant to
the equilibrium position. The different set ($k_L$, $k_R$) control
the degree of asymmetry (see Fig. \ref{fig_potential}(a) for several
examples). It returns to the classical harmonic model when $%
k_{L}=k_{R}\ne0$. If one of the constants equals zero, it
will works as a collision model. To establish our EPT theory, we need a full study of the piecewise harmoic model. 

We first study a two-particle system with the piecewise harmonic potential for the illuminating purpose.
The equations of motion with periodic boundary conditions are 
\begin{equation}  \label{eq-twoparticle-motion}
\begin{cases}
\ddot{x}_1=k_{L,R}(x_2-x_1)-k_{L,R}(x_1-x_2) \\ 
\ddot{x}_2=k_{L,R}(x_1-x_2)-k_{L,R}(x_2-x_1)%
\end{cases}%
,
\end{equation}
and we assume $x_1(0)-x_2(0)<0$ while $t=0$, then Eq.\ref%
{eq-twoparticle-motion} can be rewritten as 
\begin{equation}  \label{eq-twoparticle-motion-reW}
\begin{cases}
\ddot{x}_1=(k_R+k_L)x_2-(k_R+k_L)x_1 \\ 
\ddot{x}_2=(k_L+k_R)x_1-(k_L+k_R)x_2%
\end{cases}%
.
\end{equation}
In addition, momentum and energy are conserved, i.e., $p_{1}(t)+p_{2}(t)=0~%
\&~\frac{1}{2}p_{1}(0)^2+\frac{1}{2}p_{2}(0)^2+\frac{1}{2}%
k_L(x_{1}(0)-x_{2}(0))^2+\frac{1}{2}k_R(x_{2}(0)-x_{1}(0))^2=2\varepsilon$
(here $\varepsilon$ is mean energy of each particle). So we get $%
x_1(0)=-x_2(0)$, and $\varepsilon=\frac{1}{2}p_1(0)^2+(k_L+k_R)x_1(0)^2=%
\frac{1}{2}p_2(0)^2+(k_L+k_R)x_2(0)^2$. Then the results can been easily
obtained and the simplified form is 
\begin{equation}  \label{eq-twoparticle-result}
\begin{cases}
x_{1}(t)=\Lambda cos(2\sqrt{\bar{k}}t+\varphi) \\ 
x_{2}(t)=-\Lambda cos(2\sqrt{\bar{k}}t+\varphi)%
\end{cases}%
,
\end{equation}
where $\bar{k}=\frac{k_L+k_R}{2}$, $\Lambda=\sqrt{\frac{\varepsilon}{2\bar{k}%
}}$, $\varphi=-arctan(\frac{\sqrt{2}}{2\sqrt{\bar{k}}}\frac{p_{1}(0)}{%
x_{1}(0)})$. It's easy to see the result does not depend on the initial
conditions (if we assume $x_1(0)\ge x_2(0)$, we will get a same result).

We can clearly see the frequency of the system is $\omega=2\sqrt{\bar{k}}=2%
\sqrt{\bar{k}}sin(\frac{\pi}{2})$, which only relies on the mean value of $%
k_L$ and $k_R$ but has no direct dependency upon themselves. Besides, the
frequency is completely same as that of the pure harmonic lattice \cite%
{KittelBook}. It means that the piecewise linear two-particle system can be
strictly described by a harmonic one with force constant $\bar{k}$.

For multi-particle situation 
\begin{equation}  \label{eq_omegaC_AsyH}
\omega=2\sqrt{\bar{k}}|sin(\frac{1}{2}aq)|
\end{equation}
where $a$ is the lattice constant that is set be unit in our study, and $q$
is the wave-vector.

\begin{figure}[htbp]
\centering
\includegraphics[width=1\columnwidth]{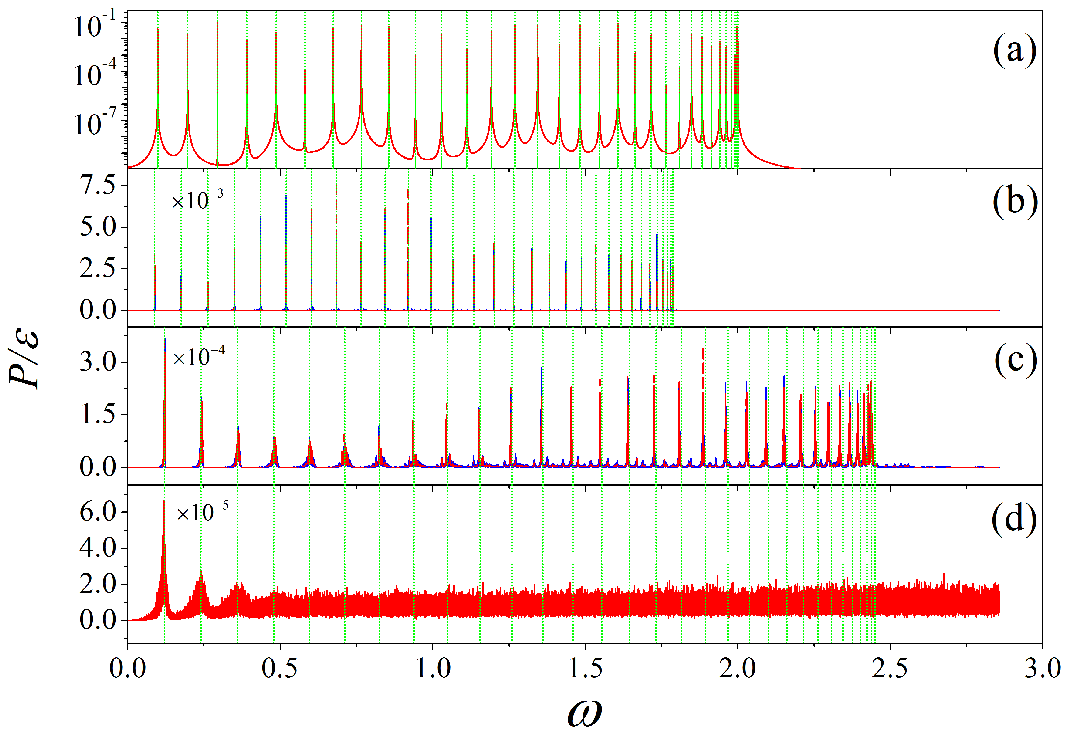}\newline
\caption{(Color lines) Normalized power spectra of the momentum time series.
(a) The red solid line is corresponding to pure harmonic chain with $%
k_L=k_R=1.0$. (b) The blue solid (or red dash) line is corresponding to $%
k_L=0.75~\&~k_R=0.85$ (or $k_L=0.85~\&~k_R=0.75$). (d) the red solid
corresponding $k_L=3.0~\&~k_R=0.0$. Energy density $\protect\varepsilon$ is
set be unit in these three simulations. (c) The blue solid line and red dash
line are, respectively, corresponding to $\protect\varepsilon=1.0,~10.0$
with fixed $k_L=1.8~\&~k_R=1.2$. The vertical green dot lines in all are
predicted centre position of frequency. System size $N=64$ for all. }
\label{fig_PowerAsyHarmonic}
\end{figure}

To verify our treatment for multi-particle cases, we immediately analyze the
power spectrum of the time series of the particles' instantaneous momentum
by fast Fourier transformation (FFT) with molecular dynamics simulation
(MDS). And the results are presented in Fig.\ref{fig_PowerAsyHarmonic}.

In Fig.\ref{fig_PowerAsyHarmonic}(a), the red solid line is power spectra of
pure harmonic chain, i.e., to set $k_L=k_R$ in our model (here we set $%
k_L=k_R=1.0$%
). The green dot line is the theoretical position of frequency. It is
clearly seen that our numerical results and existing theoretical results 
\cite{KittelBook} are consistent, namely, our calculation program is
completely accurate. Then we calculated our model and the results are shown
in Fig.\ref{fig_PowerAsyHarmonic}(b) to (d). The blue solid (or red dash)
line in Fig.\ref{fig_PowerAsyHarmonic}(b) is normalized power corresponding
to $k_L=0.75~\&~k_R=0.85$ (or $k_L=0.85~\&~k_R=0.75$), and the green dot
line is the theoretical value of frequencies' centre position that is
calculated by Eq.\ref{eq_omegaC_AsyH}. It is clearly seen that the
theoretical and numerical results fit perfectly in all frequency regime. An
interesting is that the two lines are almost complete overlap. 
This is due to exchanging the order of $k_L$ and $k_R$ does not affect the
mean value of $\bar{k}$ (the two systems are mirror symmetrical). We also
note that $\bar{k}$ does not rely on the temperature of system. The effect
can be observed in Fig.\ref{fig_PowerAsyHarmonic}(c), where we did numerical
experiments with fixed $k_L=1.8~\&~k_R=1.2$ under different energy density $%
\varepsilon=1.0,~10.0$. But the centre position of frequencies are still
identical. At the same time, we find that the normalized (by energy density)
power spectra overlap completely. This is because the model owns a similar
certain scale that is pointed in Ref.\cite{ZhongYiCPB13}. Fig.\ref%
{fig_PowerAsyHarmonic}(d) shows an extremely asymmetric case ($%
k_L=3.0~\&~k_R=0.0$) that is very like a collision model. Strictly speaking,
it is not a lattice but a fluid. But Eq.\ref{eq_omegaC_AsyH} is still able
to give a precise prediction for the centre positions of some low
frequencies (the centre position of lowest frequency is identical with $%
k_L=1.8~\&~k_R=1.2$ because they have same mean value). It states that the
piecewise linear system can be equivalent to a harmonic lattice with force
constant $\bar{k}$, seriously.

To make a comparison from (a) to (d) in Fig.\ref{fig_PowerAsyHarmonic}, we
can easily see that the phonon peak is broadening with asymmetry increasing
(see the most typical examples Eq.\ref{eq_omegaC_AsyH}(c) and(d)). A similar
phenomenon is also reported in Refs.\cite%
{ZhangyongarXiv2013,JunjiePRE2015,Zhangyong2015}. Therefore, here we
emphasize that an asymmetric piecewise linear system can be well described
by a pure harmonic lattice, only indicate that the centre positions of
phonons peaks are same as an equivalent harmonic lattice, but there is
essential difference comes from asymmetry between them, e.g., the phonon
peak will broaden in true asymmetric piecewise linear lattice, which means
that there exist stronger interactions between phonons. Besides, normal heat
conduction is also observed in asymmetric harmonic model \cite{ZhongYiCPB13}%
, but these are absent in the equivalent pure harmonic chain.

All above presented results, the system size is fixed to $N=64$, so that we
can clearly see all frequencies. But the correctness has been proved by our
numerical experiment (not shown here) that is independent with system size.

Up to now, we established our template model. For an unharmonic lattice, our goal is to obtain the
two effective force constants.
It is well known that the potential of a linear force ($F=-kx$) is $V=\frac{1%
}{2}kx^2$, which can easily obtained by $V=-\bar{F}x$ (here $\bar{F}=\frac{1%
}{2}F$ is the average force), and only the linear force has this nature.
Based on this principle, we think that if a nonlinear force can be
equivalent to a linear one, the equivalent potential should be also obtained
by the same way, i.e., $\tilde{V}_h=\langle\frac{1}{2}\tilde{k}%
x^2\rangle=\langle-\bar{F}x\rangle$, and $\bar{F}=-\frac{1}{2}\frac{\partial
V}{\partial x}$, the $\langle\cdots\rangle$ indicates the ensemble average.
Further, we can get an specific equivalent approach as 
\begin{equation}  \label{eq_KLR_FPU}
\begin{cases}
\tilde{k}_{L}=\frac{\langle \frac{\partial V}{\partial x}x\rangle}{\langle
x^2\rangle}, & x<0 \\ 
\tilde{k}_{R}=\frac{\langle \frac{\partial V}{\partial x}x\rangle}{\langle
x^2\rangle}, & x\ge0 \\ 
\tilde{k}=\frac{\tilde{k}_L+\tilde{k}_R}{2} & 
\end{cases}%
,
\end{equation}
and above equations can be further rewritten as a general integral form 
\begin{equation}  \label{eq_K_FPU}
\tilde{k}=\frac{1}{2}\big(\frac{\int_{-\infty}^{0}\frac{\partial V}{\partial
x}x\rho(x)dx}{\int_{-\infty}^{0}x^2\rho(x)dx}+\frac{\int_{0}^{\infty}\frac{%
\partial V}{\partial x}x\rho(x)dx}{\int_{0}^{\infty}x^2\rho(x)dx}\big),
\end{equation}
where $\rho(x)=e^{-\frac{V(x)+Px}{k_BT}}$ is distribution function of
relative displacement, $p=-\langle\frac{\partial V}{\partial x}\rangle$ is
the thermodynamic pressure\cite{SpohnJSP14,SpohnarXiv2015}, and $k_B$ is
Boltzmann constant that set to be unit, and $T$ is the temperature of
system. Note that $\tilde{k}$ has a certain relationship with the speed of
sound $c_s=\sqrt{\tilde{k}}$ in our dimensionless models. It is not a fixed
constant but a function of temperature $T$ and system parameters. As a
concrete example, FPU-$\alpha\beta$ model, $\tilde{k}=f(T,\alpha,\beta)$.

For symmetric FPU-$\beta$ model ($\alpha=0$) the pressure $p=0$, $V(x)$ and $%
\rho(x)$ will become even function, at this moment Eq.\ref{eq_K_FPU} can be
simplified as 
\begin{equation}  \label{eq_K_FPU_B}
\tilde{k}=\frac{\int_{-\infty}^{\infty}\frac{\partial V}{\partial x}%
x\rho(x)dx}{\int_{-\infty}^{\infty}x^2\rho(x)dx}.
\end{equation}
A interesting observation is that Eq.\ref{eq_K_FPU_B} is completely same as the form
given in Ref.\cite{BaowenEPL2006}. So our tactics at least is
effective for symmetric nonlinear models. In addition, from Eq.\ref%
{eq_K_FPU} we can see that if the pressure is known, we can get the
equivalent force constant $\tilde{k}$ through integration directly instead
of MDS. Hence, it is necessary for us to find an approach to calculate the
pressure. This goal is achived based on Spohn's recent works \cite{SpohnJSP14,SpohnarXiv2015,SpohnPRL13}. Following these works 
we develop an algorithm for calculating the pressure of a momentum-conserved lattice as 
\begin{equation}  \label{eq_pressure}
\begin{cases}
p=\frac{V(-\lambda)-V(\lambda)}{2\lambda} & (a) \\ 
\phi_{+}(\lambda)=\int_{0} ^{\infty} x e^{-\frac{V(x)+px}{T}}dx & (b) \\ 
\phi_{-}(\lambda)=\int_{0} ^{\infty} x e^{-\frac{V(-x)-px}{T}}dx & (c) \\ 
\phi(\lambda)=\phi_{+}(\lambda)-\phi_{-}(\lambda)\equiv 0 & (d)%
\end{cases}%
,
\end{equation}
where $\lambda$ is an arguments that is governed by (d). Specific steps is
to replace (a) into (b) and (c) in turn, then to solve the equation (d),
after $\lambda$ is received the pressure will be easily obtained only need
to put $\lambda$ back into (a). It is clearly seen that $\lambda$ is not
unique but arbitrary for symmetric model. Nevertheless, arbitrary nonzero
values of $\lambda$ put into the equation (a) will get fixed zero pressure,
which just agree well with expectation of symmetric model. We should point
out that if $\int_{-\infty}^{\infty}\rho(x)dx\rightarrow\infty$, the
algorithm will do not work. The internal pressure and $\tilde{k}$ can only
be acquired by MDS with Eq.\ref{eq_KLR_FPU} (e.g., FPU-$\alpha$ model and
L-J model).

\begin{figure}[htbp]
\centering
\includegraphics[width=1\columnwidth]{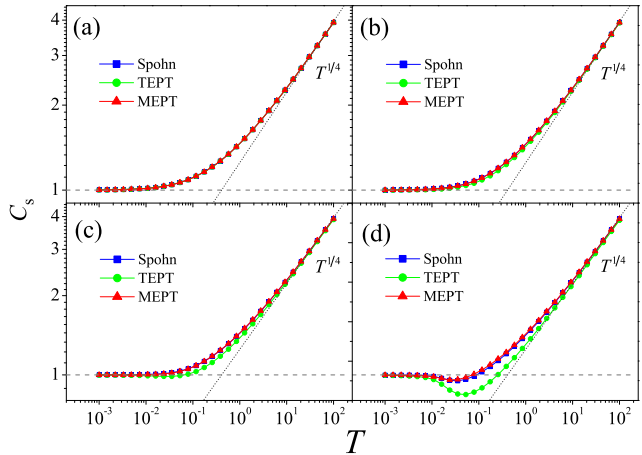}
\caption{(Color lines) The speed of sound $c_s$ against temperature $T$ with
different $\protect\alpha=0.0 (a), 1.0 (b), 1.5 (c) and 2.0 (d)$. Blue square line is calculated by Spohn's formula, green
circles are obtained by traditional EPT, and
red uptriangles are achieved by our formula. $\protect\beta=1.0$ is fixed throughly.}
\label{fig_soundSpeed}
\end{figure}

In the following, we test whether Eq.\ref{eq_K_FPU} is suitable for the
asymmetric situation by the numerical integration. Here, we only need to
check whether the prediction of the sound velocity accurately instead of
calculating the centre position of lowest frequency by FFT. This is because
Ref.\cite{ZhangyongarXiv2013} have proved that the shift of lowest frequency
and the speed of sound are equal in dimensionless models. We compare our
theoretical results (red uptriangle lines in Fig.\ref{fig_soundSpeed}) with
the standard sound velocity computed via the method developed by Spohn\cite%
{SpohnJSP14,SpohnarXiv2015,SpohnPRL13} (blue square lines in Fig.\ref%
{fig_soundSpeed}, which can also be calculated by numerical integration). In
order to make a comparison, we also show the result obtained with the traditional EPT\cite{BaowenEPL2006} (shorted as TEPT in the figure). All these are calculated by numerical integration
instead of MDS, so they are independent of system size. One can see that the
three methods get same results in low temperature regimes where the velocity
is tending to one (gray dash line) and high temperature regimes $c_s\propto
T^{1/4}$ (gray dot line, this can be also observed in Ref.\cite{HePRE2008}).
In the moderated temperature regimes, significant deviation appears with the traditional treatment, especially for large $\alpha$.
In all the regimes, our theoretical predictions show no deviations. 

\begin{figure}[htbp]
\centering
\includegraphics[width=1\columnwidth]{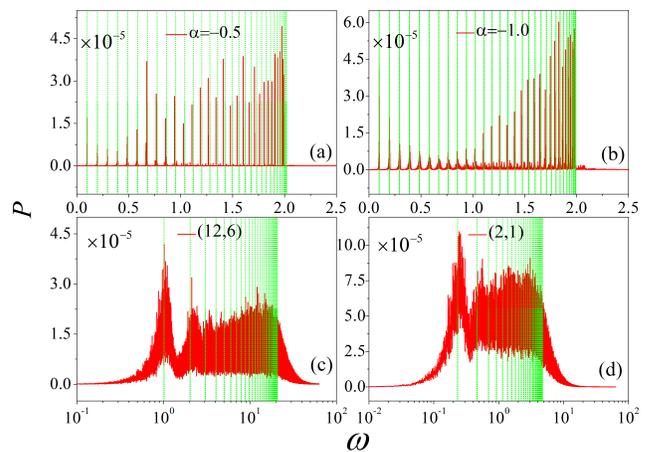}\newline
\caption{(Color lines) Power spectra of the momentum time series. (a) and
(b) for FPU-$\protect\alpha$ model, $\protect\varepsilon=0.01$. (c) and (d)
for L-J model, $\protect\varepsilon=0.5$. The vertical green dot lines are predicted centre positions of frequencies. System size is fixed at $N=64$. }
\label{fig_PowerLJ}
\end{figure}

Next, we test it in FPU-$\alpha$ model and L-J model. For FPU-$\alpha$, we
apply a small energy density $\varepsilon=0.01$ to avoid runaway instability
of trajectories. The results are drawn in Fig. \ref{fig_PowerLJ}(a) for $%
\alpha=-0.5$, and Fig. \ref{fig_PowerLJ}(b) for $\alpha=-1.0$. For L-J
model, $\varepsilon=0.5$ is fixed, and the results are shown in Fig.\ref%
{fig_PowerLJ}(c) for $(m,n)=(12,6)$, and Fig.\ref{fig_PowerLJ}(d) for $%
(m,n)=(2,1)$. From these results, we clearly see that our approach is very
workable. And phonons peaks are broadening with asymmetry increasing as well
in these models. Especially, it is clearly seen that the results of L-J
model are vary similar to the result of a fluid (compare with Fig. \ref%
{fig_PowerAsyHarmonic}(d), normal heat conduction is observed in both of them%
\cite{shunda_arXiv2012,ZhongYiCPB13} ).

To summarize, we present an effective phonon treatment that works very well in
one dimensional momentum-conserved lattice with either symmetric or
asymmetric interaction potentials. In
asymmetric models, all phonons peaks are broaden, and the degree of the broaden depends on the degree of the potential asymmerty. The more stronger the asymmetry, the more stronger the phonon scattering. This effect may explain why the normal heat conduction
behavior appears in certain lattice models \cite%
{ZhongYiCPB13,shunda_arXiv2012,ZhongYiPRE12,Savin14PRE}. We shall discuss this problem in a forthcoming paper \cite{Weicheng02}.



\end{document}